\documentclass[preprint]{aastex61}

\usepackage{apjfonts}
\received{February 2018}
\revised{March 2018}
\accepted{March 2018}
\submitjournal{ApJ}

\shorttitle{3D dust distribution}
\shortauthors{Li et al.}

\usepackage{CJK}
\usepackage[T1]{fontenc}
\begin{document}
\begin{CJK}{UTF8}{gbsn}
\title{three-dimensional structure of the milky way dust: modeling of LAMOST data}

\correspondingauthor{Shiyin Shen}
\email{ssy@shao.ac.cn}

\author{Linlin Li (李林林)}
\affiliation{Key Laboratory for Research in Galaxies and Cosmology, Shanghai Astronomical Observatory, Chinese Academy of Sciences, 80 Nandan Road, Shanghai, China, 20030}
\affiliation{University of Chinese Academy of Sciences, 19A Yuquanlu, Beijing, China, 100049}
\author{Shiyin Shen (沈世银)}
\affiliation{Key Laboratory for Research in Galaxies and Cosmology, Shanghai Astronomical Observatory, Chinese Academy of Sciences, 80 Nandan Road, Shanghai, China, 20030}
\affiliation{Key Lab for Astrophysics, Shanghai 200234 China}
\author{Jinliang Hou (侯金良)}
\affiliation{Key Laboratory for Research in Galaxies and Cosmology, Shanghai Astronomical Observatory, Chinese Academy of Sciences, 80 Nandan Road, Shanghai, China, 20030}
\affiliation{University of Chinese Academy of Sciences, 19A Yuquanlu, Beijing, China, 100049}
\author{Haibo Yuan (苑海波)}
\affiliation{Department of Astronomy, Beijing Normal University, Beijing 100875, People’s Republic of China.}
\author{Maosheng Xiang (向茂盛)}
\affiliation{National Astronomical Observatories, Chinese Academy of Sciences, Beijing 100012, People’s Republic of China.}
\author{Bingqiu Chen (陈丙秋)}
\affiliation{South-Western Institute for Astronomy Research, Yunnan University, Kunming 650500, People’s Republic of China.}

\author{Yang Huang (黄样)}
\affiliation{South-Western Institute for Astronomy Research, Yunnan University, Kunming 650500, People’s Republic of China.}
\affiliation{Department of Astronomy, Peking University, Beijing 100871, P. R.}
\affiliation{Kavli Institute for Astronomy and Astrophysics, Peking University, Beijing 100871, P. R. China.}
\author{Xiaowei Liu (刘晓为)}
\affiliation{South-Western Institute for Astronomy Research, Yunnan University, Kunming 650500, People’s Republic of China.}
\affiliation{Department of Astronomy, Peking University, Beijing 100871, P. R.}
\affiliation{Kavli Institute for Astronomy and Astrophysics, Peking University, Beijing 100871, P. R. China.}



\begin{abstract}
We present a three-dimensional modeling of the Milky Way dust distribution by fitting the value-added star catalog of LAMOST spectral survey. The global dust distribution can be described by an exponential disk with scale-length of 3,192 pc and scale height of 103 pc. In this modeling, the Sun is located above the dust disk with a vertical distance of 23 pc. Besides the global smooth structure, two substructures around the solar position  are also identified. The one located at  $150^{\circ}<l<200^{\circ}$ and $-5^{\circ}<b<-30^{\circ}$ is consistent with the Gould Belt model of \citet{Gontcharov2009}, and the other one located at $140^{\circ}<l<165^{\circ}$ and $0^{\circ}<b<15^{\circ}$ is associated with the Camelopardalis molecular clouds.

\end{abstract}




\section{Introduction} \label{sec:intro}

Dust makes up just about $1\%$ of the interstellar medium (ISM), but plays important role in a number of physical and chemical processes.
Dust is mainly created by stars, either in the atmospheres of AGB stars \citep{Indebetouw2014,Dwek2011} or during supernova explosions \citep{Ferrarotti2006}. On the other hand, dust is also thought to be a catalyst for the production of molecular hydrogen \citep{Hollenbach1971} and consequently be connected with star formation \citep{Bigiel2008,Casasola2015,Azeez2016}. The spatial distribution of dust in galaxies not only provides a bridge to the formation of stars  but also contains information of the cycling of metals among gases and stars.

Besides the findings of the strong spatial correlations between the surface mass densities of dust and molecular hydrogen \citep{Foyle2012,Pappalardo2012,Hughes2014}, the comparison of the dust and stellar distribution has also been done for many  nearby spiral galaxies. These works suggested that the dust disk tends to be larger radially while thinner in the vertical direction than the stellar disk. For example, \cite{Bianchi2007} analyzed a sample of seven nearby edge-on galaxies observed in the V and K -bands, showing that the ratio of scale-length of dust disk $h_d$ to stars $h_s$ is about 1.5 and  scale-height ratio $z_d /z_s$ is about 1/3; \cite{Geyter2014} investigated 12 edge-on galaxies and found that the dust disk is about 75\% more radially extended but only half as high as the stellar disk. More recently, \cite{Casasola2017} studied the radial distribution of dust and stars in DustPedia \citep{Davies2017}  face-on galaxies and found that  the dust-surface-density scale-length is about 1.8 times the stellar density, in agreement with edge-on galaxies.

Unlike the extra-galactic galaxies, the structural study of our Milky Way takes advantages of resolving individual stars while also suffers the disadvantages of excessive details. A detailed review of the stellar structural parameters of the Milky Way can be found in \citet{Bland2016} . For the stellar thin disk, the general conclusion is that it can be fitted by an exponential disk with scale-length $2.6 \pm 0.5$ kpc and scale-height $220 \sim 450$ pc.  For the dust component, it has been shown that, despite of  spiral arms, flares and warps \citep{Drimmel2001,Marshall2006,Amres2005,Reyle2009}, the global distribution of dust also follows an exponential disk \citep{Misiriotis2006,Jones2011}. However, the scale-length and scale-height of the dust disk have not been constrained very well yet.

\cite{Drimmel2001} presented a three-dimensional Galactic dust distribution model and obtained a scale-length of 2.26 kpc and scale-height of 134.4 pc by fitting the FIR data from COBE/DIRBE, while \cite{Misiriotis2006} got the results of  5.0 kpc and 100 pc respectively also by modeling FIR emission of  COBE data.
\cite{Jones2011} created a three-dimensional Galactic extinction map using the spectra of more than 56,000 M dwarfs in SDSS (Sloan Digital Sky Survey, \citealt{York2000}) and then obtianed the dust scale-height  of 119 pc around solar region. To get a more reliable modeling of Galactic dust distribution, a larger sample of dust extinction tracer is better. The LAMOST (Large Sky Area Multi-Object Fiber Spectroscopic Telescope) spectral survey \citep{Cui2012,Zhao2012} has obtained the largest spectroscopic sample of Galactic stars to date (more than eight million). By applying the standard pair technique \citep{Yuan2014} on the LAMOST and SDSS spectra, the distance and dust extinction of millions of stars have been estimated. With the detailed three-dimensional Galactic extinction map centered at solar position, the global Galactic dust distribution may also be outlined.


In this work, we aims at constraining the scale-length and height of the Galactic dust distribution to an unprecedented statistical accuracy  with  LAMOST data. The outline of this paper is as follows. In Section \ref{sec:data}, we introduce the data set used in modeling in more detail. We present the dust distribution model and fitting method in Section \ref{sec:highlight}. Our main results are shown and discussed in Section \ref{sec:result}. Finally, we present a brief summary in Section \ref{sec:conc}.

\section{data} \label{sec:data}
\subsection{dust extinction of stars}\label{sec2.1}

Using the stellar parameters estimated from LAMOST Stellar parameter Pipeline at Peking University (LSP3, \citealt{Xiang2015})
 and applying the standard pair technique, Yuan et al. obtained the Galactic extinction  to a sample of  $\sim 6$ million stars from LAMOST spectral survey and SDSS data release 9. In specific, after pairing the stars with the same spectral type but different Galactic extinction, the reddening in different colors are calculated using the the photometry from GALEX ($FUV,NUV$), SDSS ($u,g,r,i,z$), XSTPS-GAC/APASS ($g,r,i$), 2MASS ($J,H,Ks$) and WISE ($W1,W2,W3,W4$) bands. These reddening in different colors are then converted to the standard reddening $E(B-V)$ using the extinction coefficients of \cite{Yuan2013}. The final $E(B-V)$ values are then the weighted mean of the results from different colors. The typical uncertainty of final $E(B-V)$ of stars is  about 0.04 mag. With $E(B-V)$ of stars, their distances  are then calculated from their  photometric absolute magnitudes. For stars with high-quality spectra, the absolute magnitudes are estimated from stellar parameters $T_{\rm{eff}}$ (effective temperature), log $g$ (surface gravity) and [Fe/H] (metal abundance) \citep{Yuan2015}. Alternatively, distance of stars with low $S/N$ spectra (22\% with $S/N<10$) are calculated using the color-magnitude relation of \cite{Ivezi2008}. Depending on the method and data quality, the uncertainty of the distance estimation  varies  from 10 to 30 percent.

\subsection{average extinction of grids}\label{sec2.2}

Considering the small Galactic extinction values  at high Galactic regions,  we only use the stars in the region $|b|\leqslant30^{\circ}$. The number of stars in the resulted sample is 4,367,136.  However, the dust extinction of individual stars is not an ideal tracer of the global distribution of Galactic diffuse dust, because of its large uncertainty and being susceptible to the sub-structures (e.g. molecular clouds) along the line-of-sight. To alleviate this effect, we group the stars into small volume grids and calculate their average Galactic extinction  in each grid. We set the width ($\bigtriangleup l$) and length ($\bigtriangleup b$) of each grid to be $2.5^{\circ}$.  In terms of depth, we set $\bigtriangleup d = 250$ pc to ensure that it is larger than a single molecular cloud (even for giant molecular clouds, \citealt{Murray2011}). As the observed star number density decreases with increased distance $d$,  we set $\bigtriangleup d=500$ pc when $d>2.5$ kpc. Finally, we obtain 26,363 grids.

For each grid,   we take the median $E(B-V)$ of the stars  as the reddening of the grid and noted as $E_{obs}$, while the coordinate ($l_{i},b_{i},d_{i}$) is represented by the mean  positions of the stars.

\begin{figure}[!htp]
\epsscale{1}
\centering
\includegraphics[scale=1]{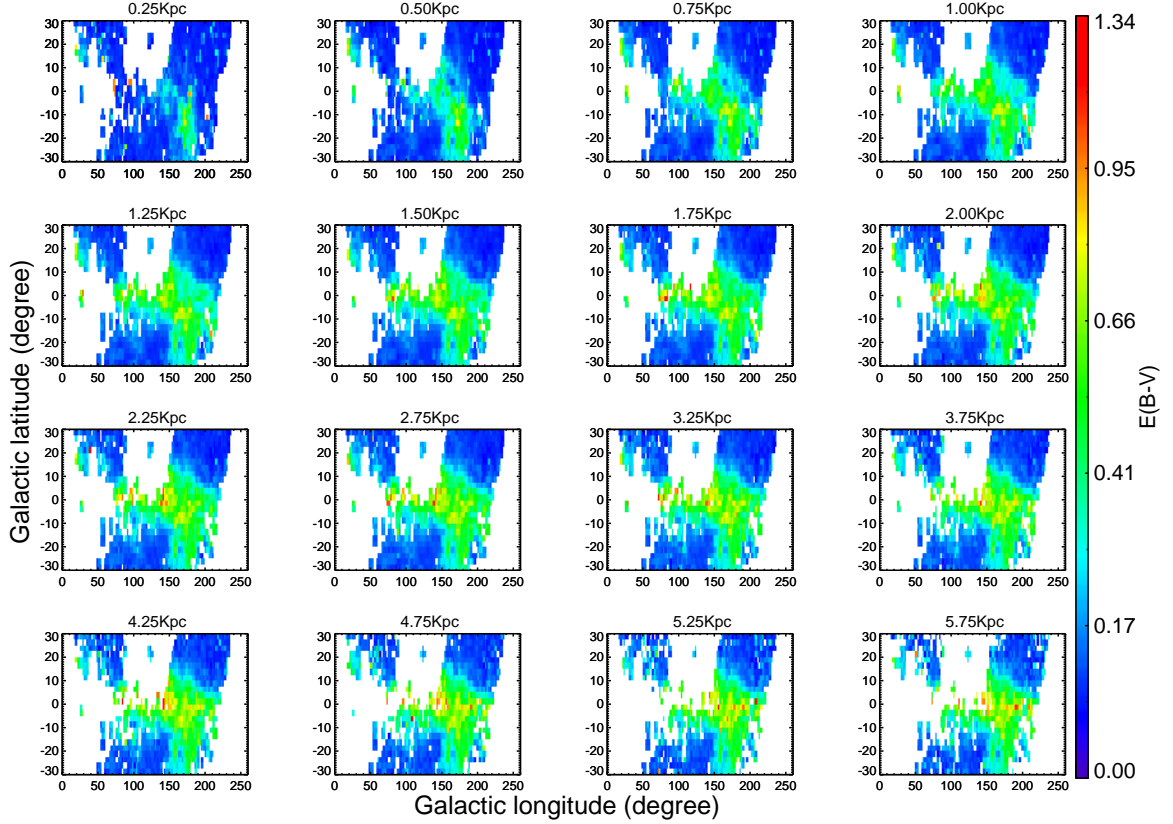}
\caption{The Galactic dust reddening map at different distance intervals.  The x and y axis are Galactic longitude and Galactic latitude, respectively.
\label{diskebv}}
\end{figure}

 In Figure \ref{diskebv}, we show the images of $E_{obs}(l_{i},b_{i},d_{i})$  at distance intervals of 250 pc for $d<2.5$ kpc and 500 pc for $d>2.5$ kpc. As can be seen, $E_{obs}$ increases monotonically and significantly  from $d=0.25$  to 2.75 kpc, especially in the low Galactic latitude region.

\section{method} \label{sec:highlight}
\subsection{model}\label{sec3.1}

In optical wavelengths, the Galactic dust extinction at a given wavelength can be parameterized by a simple screen model,
\begin{equation} \label{eq1}
 A_{\lambda} = 1.086\tau_{\lambda},
\end{equation}
where $\tau_{\lambda}$ is the optical depth.
If  the physical properties of  dust grains in Galactic ISM are uniform,  $\tau_\lambda$ is then proportional to the the dust column density along the line-of-sight,

\begin{equation} \label{eq2}
 \tau_{\lambda}(s)=\int_{0}^{s}\rho(s)\sigma_{\lambda} ds,
\end{equation}
where $\rho$ and $\sigma_{\lambda}$ are the density and effective cross section (at wavelength $\lambda$) of dust grains respectively. Physically, $\sigma$ is a function of $\lambda$ and is determined by  the chemical composition of the dust grains. In observation, this function can be parameterized by the dust extinction curve $k_{\lambda}$,
\begin{equation} \label{eq3}
\kappa_{\lambda}=A_{\lambda}/E(B-V) \,.
\end{equation}

 Observations show that there is variation of the extinction curve in different Galactic regions \citep{Savage1979,Mathis1981,Cardelli1988} , which
is typically parameterized by different $R_V$ values \citep{Cardelli1989}.  However, this variation is not very significant on large scales and a standard Galactic extinction curve has $R_V=3.1$ \citep{Draine2003}. Recently, using data from  APOGEE spectroscopic survey and Pan-STARRS1 photometry, \cite{Schlafly2016} shows that the  $R_V$ distribution has  a mean value 3.32 and  dispersion $\sigma(R_V)$=0.18. 

Giving the  Galactic dust distribution model and assuming the uniform chemical properties of dust grains,
the $E(B-V)$ values of any positions  relative to  the location of the Sun can be easily derived. We take the Galactocentric cylindrical coordinate system $(R,\Phi,Z)$ to characterize the Galactic dust distribution, where $R$ is  the Galactocentric radius, $\Phi$ is the position angle and $Z$ is the vertical distance. We assume that the dust distribution is axisymmetric and can be parameterized by an exponential disk \citep{Misiriotis2006,Jones2011} 

\begin{equation} \label{eq4}
\rho_{d}(R,\Phi,Z)=\rho_0exp(-R/l_0-Z/h_0)
\end{equation}
where  $\rho_0$ is the dust grain density at Galactic center, $l_0$ and $h_0$ are the scale-length and height of the dust disk respectively.

To model  $E(B-V)$ of an observed position $(l,b,d)$ relative to the Sun, we need to set the Sun's position ($R_\odot,\Phi_\odot,Z_\odot$) in the Galactocentric coordinate system. We set $\Phi_{\odot}=180^{\circ}$ and take $R_{\odot}=8.2$ kpc \citep{Bland2016}. For $Z_{\odot}$, we set it to be a free parameter.
The  coordinate conversion then can be  written as

\begin{equation} \label{eq5}
\left\{ 
\begin{array}{c}
 R=\sqrt{(d \rm{cos}(\emph{b})\rm{cos}(\emph{l})-\emph{R}_{\odot})^2+(\emph{d} cos(\emph{b})sin (\emph{l}))^2},\\
 \Phi=\arctan(\rm{cos}(\emph{b})sin (\emph{l})/(\rm{cos}(\emph{b})cos(\emph{l})-\emph{R}_{\odot}) ),\\
 Z=d \rm{sin}(\emph{b})+\emph{Z}_{\odot}.
\end{array}
\right. 
\end{equation}
With  Equ. \ref{eq4} and \ref{eq5}, we calculate the model dust reddening at any position $E_{mod}(l,b,d)$ and compare them with the observed $E(B-V)(l_i,b_i,d_i)$ values (those shown in Figure \ref{diskebv}) and then make best estimations of the Galactic structural parameters $(l_0,h_0)$, normalization parameter $\rho_0\sigma$ and Sun's vertical position $Z_{\odot}$.

\subsection{fitting procedure}\label{sec3.2}

In Section \ref{sec:data}, we have  obtained  $E_{obs}(l_{i},b_{i},d_{i})$ for 26,363 grids. With these data, we fit the above four model parameters by minimizing   

\begin{equation} \label{eq5.1}
\chi^2 (l_0,h_0, \rho_0\sigma, Z_{\odot}) = \sum_i \frac{ (E_{obs}^i-E_{mod}^i(l_0,h_0,\rho_0\sigma, Z_{\odot}) )^2}{err_i^2},
\end{equation}

where $err_i$ is the uncertainty of $E_{obs}^i$ and is computed from  bootstrapping of the sample stars in each grid. In order to have a good estimation of the uncertainties of the fitting parameters, we convert $\chi^2$ to  probability through $\chi^2=-2\rm{ln}L$ and use MCMC technique \citep{Metropolis1953,Hastings1970} to sample the probability distribution.
The model is initially fitted using all the $E_{obs}$ values, while we find some grids deviate from the smooth disk structure significantly.
This deviation is caused by the substructures of the dust distribution (see Section \ref{subs})  and would bias the fitted parameters of the smooth disk.

Similarly to \cite{Chen2017}, we remove the outlying grids (substructures) and iterate the fitting procedure. In specific, after each fitting step, the outlying grids are identified according to the fitting residuals $\epsilon=(E_{obs}-E_{mod})/E_{mod}$, where the grids with $\epsilon > \epsilon_{crit}$ are masked as outliers (substructures). The critical value $\epsilon_{crit}$ used  for defining outlying grids in each step are 1.5, 1.0, 0.7, 0.5, and reject 1955, 3601, 5458, 9132 grids, respectively.
Finally our model-fitting algorithm converges towards the smooth structure of the dust disk. That is to say, all the remained 17,231 grids are fitted well by the exponential profile with all their relative fitting residuals $\epsilon<0.5$.

\section{results and discussion} \label{sec:result}
\subsection{Structural parameters of the Galactic dust disk}

\begin{figure}[!htp]
\epsscale{1}
\centering
\includegraphics[scale=0.8]{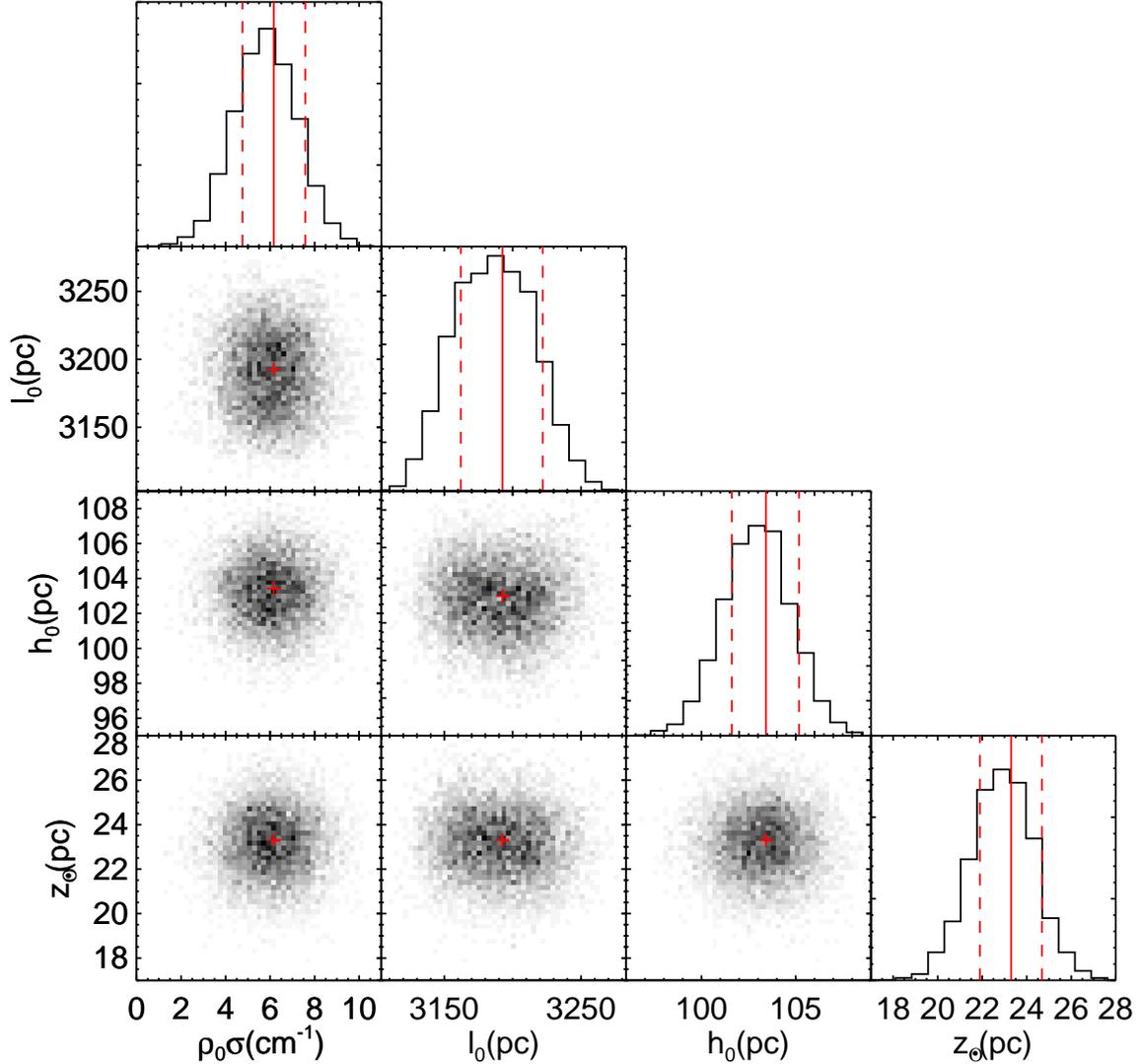}
\caption{Two-dimensional marginalized PDFs of the model parameters, $\rho_{0}\sigma, l_{0}, h_{0}$ and $Z_{\odot}$ obtained from MCMC analysis. Histograms on the top of each column show the one-dimensional marginalized PDFs of each parameter labeled at the bottom of the column. Red pluses and solid lines indicate the best solutions. The dashed lines give the 16th and 84th percentiles and are used to denote the fitting uncertainties.
\label{exclude}}
\end{figure}

The final two-dimensional probability distribution functions (PDFs) of the four fitting parameters are shown in Figure \ref{exclude}. The best-fitting parameters are $\rho_{0}\sigma=6.2\pm^{1.4}_{1.4}$ kpc$^{-1}$, $l_{0}=3,192\pm^{29}_{30}$ pc, $h_{0}=103.4\pm^{1.7}_{1.8}$ pc and $Z_{\odot}=23.3\pm^{1.3}_{1.4}$ pc. To show the goodness of our best model fitting, we show the observed and model predicted $E(B-V)$ values along the line-of-sight for four selected regions. The dots with errorbars show the median $E(B-V)$ values of the grids at different heliocentric distances, while the solid lines show the predictions from the best model. The red dots are those with fitting residuals too large ($\epsilon>0.5$) to be included in the final model fitting (see Section \ref{sec3.2}). As can be seen, the best fitting model shows good consistence with the observed $E(B-V)$ values in different directions, especially on the global trend of the $E(B-V)$ values as function of distance.

\begin{figure}[!htp]
\epsscale{1}
\centering
\includegraphics[scale=1.2]{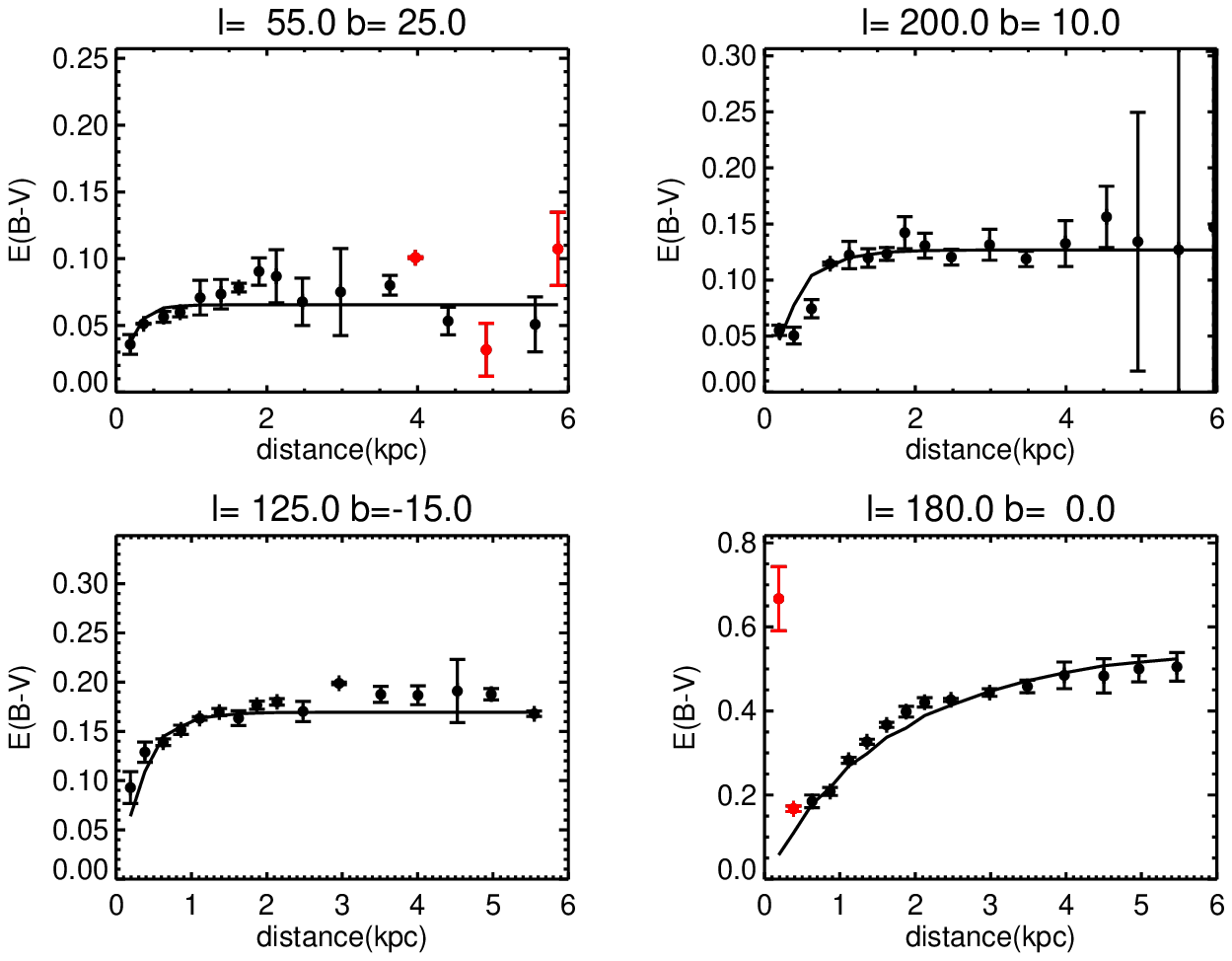}
\caption{The observed and modelled $E(B-V)$ values as function of distance in four different selected line-of-sights. The dots with errorbars are the observed values, while the solid lines are model predictions. These red dots are the outlying data with relative fitting residual larger than 50 percent.
\label{goodfit}}
\end{figure}

We list the best-fitting model parameters of the dust disk in Table \ref{tab1} and make comparisons with the structural parameters of various Galactic disk components in other studies. For the dust component, thanks to the large data set, our results show the smallest statistical scatter. The disk scale-height in different studies are in good consistences except slight lager value of \cite{Drimmel2001}. Our scale-length is larger than \cite{Drimmel2001} but much smaller than \cite{Misiriotis2006}. Both \cite{Drimmel2001} and \cite{Misiriotis2006} used the two-dimensional FIR emission to constrain the dust structure, which would have larger intrinsic degeneracy between the disk scale-length and height than our three-dimensional data. In such a modeling, as the results of \cite{Drimmel2001} and \cite{Misiriotis2006} shown, a larger disk scale-length is degenerated with smaller scale-height. Moreover, in both studies of \cite{Drimmel2001} and \cite{Misiriotis2006}, the FIR emission from the dust in molecular clouds in solar neighborhood has not been taken into account, the resulted dust structural parameters therefore might also be biased. In addition, only our modeling takes vertical position of the solar position $Z_\odot$ into account, and our model result is in excellent agreement with the stellar disk result of \cite{Ivezi2008}.

\begin{table}[!htp]
 \centering
  \caption{The structural parameters of various Galactic disk components in different studies.}
  \begin{tabular}{lcccc}
  \hline
  \hline
reference& component  & scale-length  & scale-height  & $Z_{\odot}$ \\
         & &(pc)&(pc)&(pc)\\
  \hline
This study & dust & $3192\pm^{29}_{30}$ & $103.4\pm^{1.7}_{1.8}$ & $23.3\pm^{1.3}_{1.4}$  \\
\cite{Drimmel2001}& dust  &  $2260\pm 160$& $134.4\pm 8.5$ & -- \\
\cite{Misiriotis2006} & dust & 5000 & 100 & --\\
\cite{Jones2011}& dust & -- & $119\pm 15$ & --\\
\cite{Ivezi2008} & steller & $2150\pm 400$ & $245\pm 50$ & $25\pm 5$\\
\cite{Chen2017} & steller & $2343\pm 400$ & $322\pm 30$ & --\\
\cite{Heyer2015}  & H$_2$  & --& $90 \sim 120$&--\\
\cite{Tielens2010}& H$_2$  & 3000& --& --\\
\cite{Marasco2017}& H$_2$  & & $64 \pm 12$& --\\
\cite{Kalberla2009}& HI    &  $\sim 3750$ &--& --\\
\cite{Marasco2017}& HI    & --  & $202 \pm 28$ &--\\
   \hline
\end{tabular}\\
\label{tab1}
\end{table}

\subsection{Comparisons with the structural parameters of other Galactic disk components}
We listed two groups of structural parameters of the stellar thin disk in Table \ref{tab1}, \cite{Ivezi2008} and \cite{Chen2017}. \cite{Ivezi2008} mapped the stellar disk structure using the SDSS photometric data, while \cite{Chen2017} used the data set from XSTPC-GAC and SDSS photometric surveys and cover similar region with us. \cite{Ivezi2008} also took the vertical solar position relative to the Galactic plane into account, while \cite{Chen2017} did not. Despite the differences in these details, their results of the disk scale-length and scale-height are consistent with each other inside errors. Comparing with the stellar structural parameters, our results of the dust disk show significantly larger scale-length and smaller scale-height. That means the Galactic dust disk is thinner and more radially extended than the stellar disk. This result is qualitatively consistent with the general impressions of other edge-on spiral galaxies.
 
Quantitatively, comparing with the thin stellar disk parameters of \cite{Chen2017}, our scale-length of dust disk is larger with a ratio $h_d/h_s\sim 1.4$ while the scale-height is smaller with a ratio $z_d/z_s\sim 1/3$. These ratios are in good agreement with the ratios of nearby edge-on spiral galaxies in the study of \cite{Bianchi2007} derived from  optical images. Using the SED fitting  of multi-wavelength data (from UV to far infrared) on a sample of 75 galaxies in the Spitzer Infrared Nearby Galaxies Survey (SINGS),  \cite{Mateos2009} got that the median $h_d/h_s \sim 1.1$, which is also marginally consistent with the optical only studies. However, more recently, also using SED fitting of multi-wavelength data, \citet{Casasola2017} found a much larger $h_d/h_s$  ($\sim 1.8$)  for a sample of 18 face-on galaxies in DustPedia database.  The  larger  $h_d/h_s$ values in \citet{Casasola2017}  might be caused by its inclusion of Herschel data in longer wave-lengths. As shown by \cite{Alton1998} (see also
\citealt{Davies1999}), the cold dust ($\sim$20 K) in galaxies is more extended than warm dust component ($\sim$30 K). 

For the Galactic gas component, observations show that both of the atomic and molecular gas distribution have a hole within 4 kpc of the center \citep{Tielens2010,Marasco2017}, similar  to those shown by some nearby  galaxies \citep{Casasola2007,Casasola2008,Nieten2006}. The depression of gas in Galactic center is most likely caused by the Galactic bar. Molecular gas is more closely confined to the Galactic plane than the atomic component. The surface density of the molecular gas is larger than that of HI within the solar circle, while the atomic gas dominates the molecular gas in the outer Galaxy. The different spatial distribution of atomic and molecular gas is caused by the transformation of HI to H$_2$, which is physically connected with dust particles. Therefore, it is very instructive to make a detailed comparison of the  spatial distribution of the dust component in our study with the gas component in the literature.

For Galactic HI distribution, the review of \cite{Kalberla2009} shows that at Galactocentric radii $R>R_{\odot}$ the HI surface density can be approximated by an exponential distribution with a radial scale-length of 3.75 kpc, while the scale-height increases with Galacticocentric radius and reaches 150 pc in the solar local. Recently, \cite{Marasco2017} investigated the detailed distribution of atomic hydrogen inside the solar circle, found that the HI distribution has a scale-height of $202 \pm 28$ pc. Although the discrepancy  in the values of the HI scale-height, the common conclusion is that the Galactic HI distribution is more extended than the dust in both radial and perpendicular direction.
 
For Galactic H$_2$ distribution, at radii larger than 4.5 kpc, it can also be approximated by an exponential distribution with a scale-length of about 3 kpc \citep{Tielens2010}.  The distribution of H$_2$ perpendicular to the Galactic plane shows a slight rise in scale-height, from $\sim$ 90 pc at $R = 2$ kpc to $\sim$ 120 pc at $R = 8$ kpc \citep{Heyer2015}.  The work of \cite{Marasco2017}  obtained  $64 \pm 12$ pc for the scale-height of H$_2$ traced by CO within the solar circle. As can be seen, the Galactic molecular gas have  similar values of both scale-length and scale-height as the dust distribution in our study.
 
In summary, the distributions of different components in the Galaxy indicate that dust is more related to molecular hydrogen than atomic hydrogen. This result is consistent with many earlier findings for both the Milky Way and other nearby galaxies (e.g. \citealt{Pineda2008,Bendo2010,Foyle2012, Lee2014,LeeMNRAS}), which is physically explained by the idea that dust grains  act as an efficient catalyst of atomic hydrogen reactions and can further shield molecules from the photo-dissociating radiation.  However, \cite{Casasola2017} found that the scale-length ratio between dust and molecular gas (H$_2$ derived from CO $J=1\rightarrow$0 and $2\rightarrow1$) is about 2.3 for nearby, face-on spiral galaxies. They conclude that the much steeper H$_2$ radial profile in their study can be explained by a change in the typical lifetimes of grains against destruction by shocks, with a longer lifetime at larger radii.  As  discussed in \cite{Casasola2017}, the different conclusions reached on the H$_2$-to-dust distribution may partly attributed to the different  H$_2$ tracers (different CO transition lines)  used in different studies.

 \subsection{substructures} \label{subs}

\begin{figure}[!htp]
\epsscale{1}
\centering
\includegraphics[scale=1]{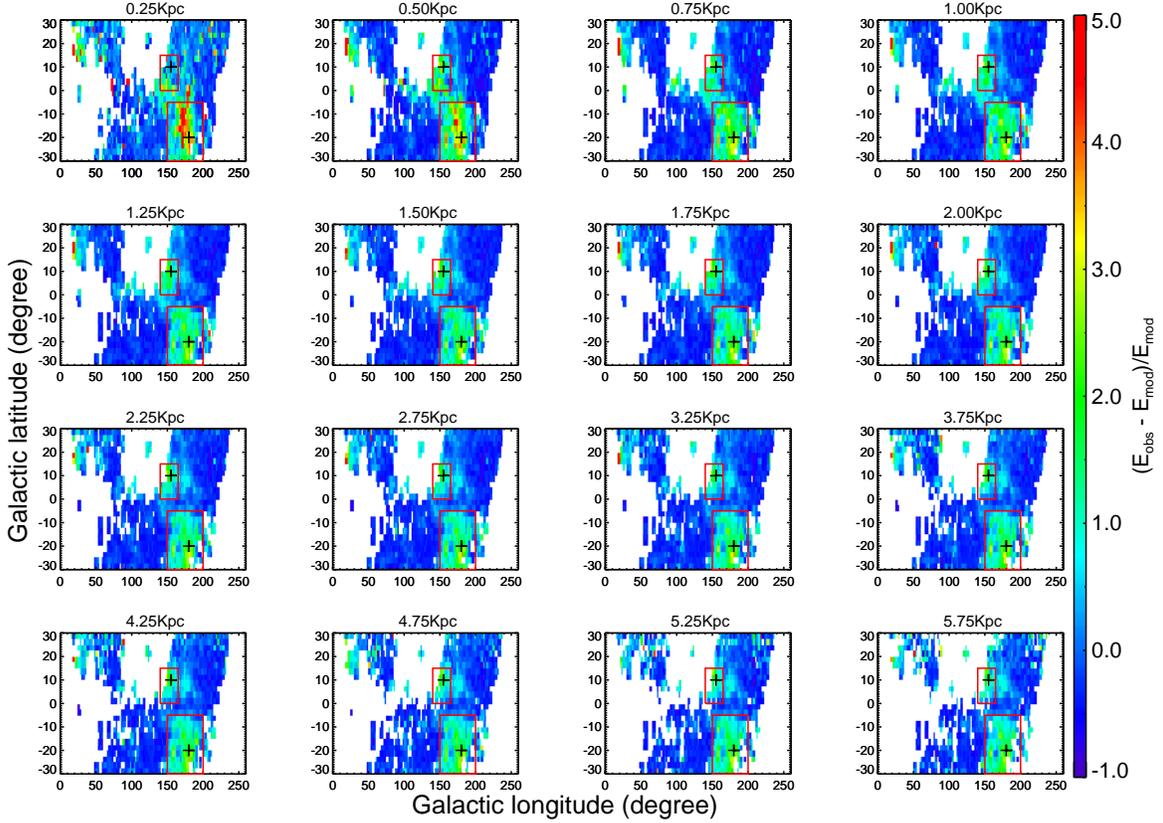}
\caption{The residual ($\epsilon=(E_{obs}-E_{mod})/E_{mod}$) maps of the best model fitting at different distance intervals. The two red boxes indicate the substructure regions with $\epsilon>0.5$. The Region A locates at $150^{\circ}<l<200^{\circ}$ and $-30^{\circ}<b<-5^{\circ}$, with the area about 1180 degree$^2$. The Region B locates at $140^{\circ}<l<165^{\circ}$ and $0^{\circ}<b<15^{\circ}$, with the area about 370 degree$^2$.
\label{residual}}
\end{figure}

As we have mentioned in Section \ref{sec3.2}, there are 9,132 outlying grids ($\epsilon >0.5$) have been masked from the best model fitting. We show the residual map $\epsilon=(E_{obs}-E_{mod})/E_{mod}$ in different distance slices of our best model fitting in Figure \ref{residual}, where the green area with $\epsilon > 0.5$ are the regions masked. As can be seen, these outlying grids are mainly located in two consecutive area, a large one at $150^{\circ}<l<200^{\circ}$ and $-30^{\circ}<b<-5^{\circ}$ (hereafter region A) and a small one at $140^{\circ}<l<165^{\circ}$ and $0^{\circ}<b<15^{\circ}$ (hereafter region B), which have been marked as two red boxes in each panel of Figure \ref{residual}. These deviations begin to appear in the first distance slice and then keeps up to the largest distance we probe. This systematical deviation is caused by the incremental effect of the dust extinction and implies dust substructures in the solar neighborhood.

\begin{figure}[!htp]
\epsscale{1}
\centering
\includegraphics[scale=1]{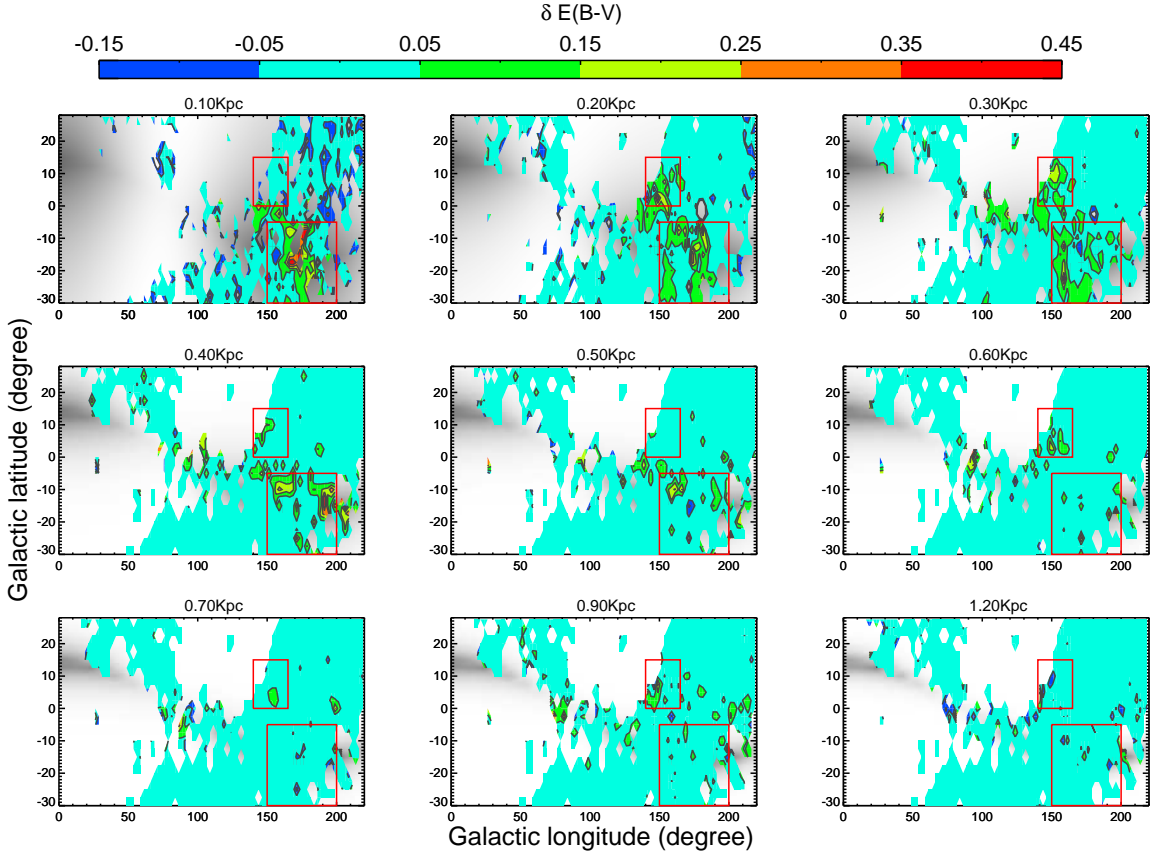}
\caption{The differential residual of dust model fitting $\delta E(B-V)$ at different distance intervals. The gray background shows the prediction from the Gould Belt model of G09. The two red boxes are the same with Figure \ref{residual}.
\label{belt}}
\end{figure}

To have a better view of the possible substructures, we redefine 9 distance bins at $d<1.2$ kpc and take a differential quantity $\delta E(B-V)$ to characterize the local deviation of $E(B-V)$ from exponential disk model prediction at each distance bin. Specifically, $\delta E(B-V)$ is defined as  $\delta E(B-V)=(E_{obs}-E_{mod})_{d_{i+1}}-(E_{obs}-E_{mod})_{d_{i}}$ where $d_{i}$ means $i$-th distance bin. Figure \ref{belt} shows how $\delta E(B-V)$ varies in different distance slices. As can be seen, for both region A and B, the dust substructure mainly located at solar neighborhood $d\sim 200-300$ pc. There are slight differences in radial extension of the two substructures. The extension of region A is about from 0 to 500 pc, Region B mainly  distributes from 200 to 300 pc and also show  structures at 600 and 900 pc  distance intervals respectively.

\begin{figure}[!htp]
\epsscale{1}
\centering
\includegraphics[scale=1.2]{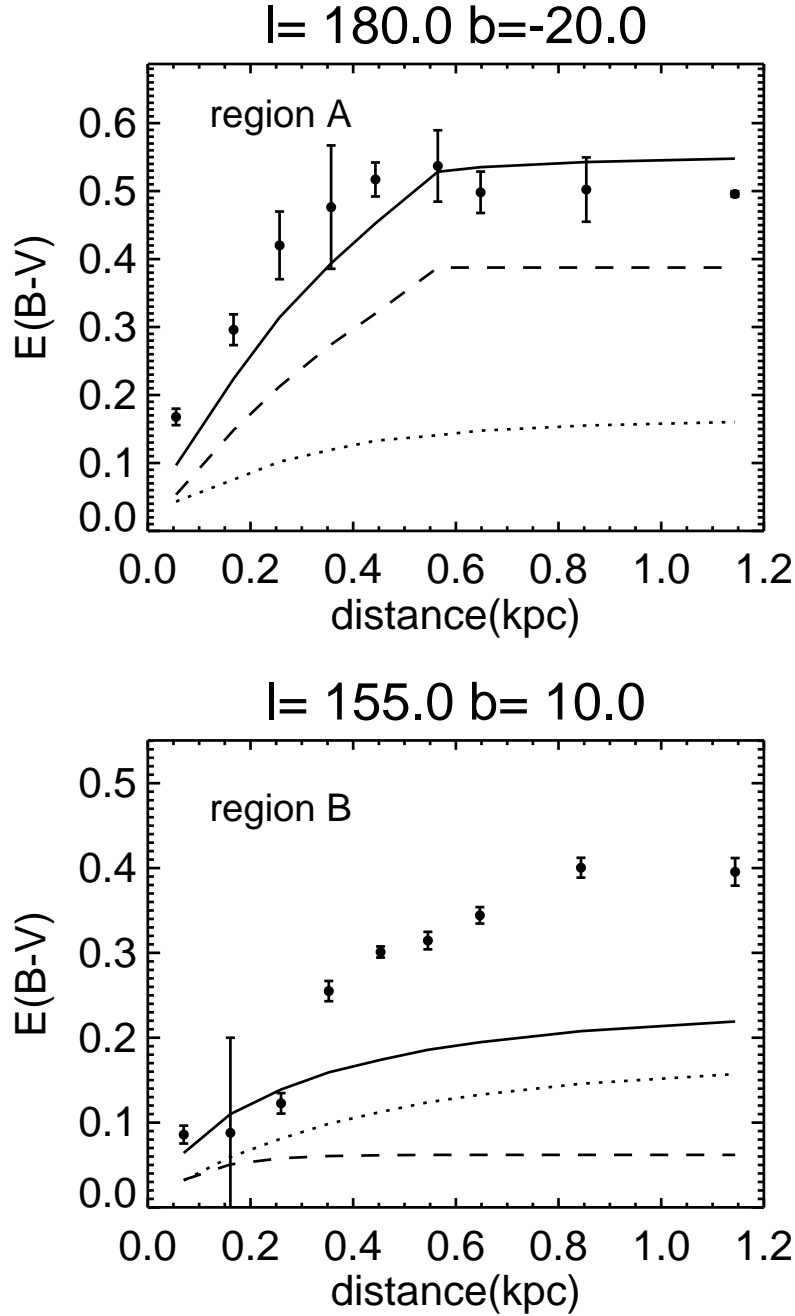}
\caption{The growth curve of the Galactic extinction values of two selected regions. The dotted, dashed and solid lines are the exponential disk, G09 and `smooth disk + G09', respectively.
\label{badfit}}
\end{figure}

At solar neighborhood, there is a known dust substructure, the Gould belt, which is a quite flat system consisting of many molecular clouds with the total gas mass about $7\times10^5 M_{\odot}$ within 500 pc. The detailed structure of the Gould belt can be found in the review paper of \cite{Bobylev2014}. \cite{Gontcharov2009} (hereafter G09) studied the interstellar extinction of Gould belt and presented an analytical three-dimensional extinction model within 500 pc of the solar position in Galactic coordinates.

To compare the Gould belt model with the two substructures we have identified, we also plot the Gould belt extinction values at each distance interval from G09 model as the gray background in each panel of Figure \ref{belt}. As can be seen, the location of substructure A is in good consistence with G09 model, while the substructure B is not. To have a more quantitative comparison, we plot the observed $E(B-V)$ growth curves of these two substructure regions and  compare them with the predictions from a `smooth disk + G09' dust structure model. 

Figure \ref{badfit} shows the results of two exampled sight-lines for two substructure areas respectively. In this plot, the dots with error-bars show the observed median $E(B-V)$ values as function of the increased distance with a step of 100 pc. The dotted lines show the predicted $E(B-V)$ values from the best exponential disk model, while the dashed lines show the $E(B-V)$ values of Gould belt from G09 model. By combining the exponential disk and G09 model, we see that the resulted $E(B-V)$ growth curve (solid line) is in good consistence with the observations for the exampled sight-line in region A (top panel). While for region B, the `smooth disk + G09' model still deviates from the observation significantly.

Region B is located inside Camelopardalis segment of the Milky way ($130^{\circ}<l<160^{\circ}$,$-12^{\circ}<b<12^{\circ}$), where many dust and molecular clouds within 1 kpc of solar position have also been identified \citep{Obayashi1999,Zdanavicius2002,Zdanavicius2005}. Studies show that the interstellar extinction at the directions of  the Camelopardalis molecular clouds  rises at $100 \sim 300$ pc and $800 \sim 900$ pc, and then reaches about $E(B-V)\approx0.4 \sim 0.8$ mag at 1 kpc. For the average dust extinction of region B we obtained, as we can see from Figure \ref{belt}, $\delta E(B-V)$  shows evident  structures at the distance intervals of 200-300, 600 and 900 pc respectively,  which is in good consistence with the extinction rises found in the literature. Therefore, we conclude that the substructure, region B, is associated with the  Camelopardalis molecular clouds.


\section{Summary} \label{sec:conc}

In this paper, we have studied the smooth structure of the dust distribution using the three-dimensional dust reddening catalog from LAMOST spectral survey. The smooth dust distribution can be fitted by an exponential disk with the scale- length of 3,192 pc and scale-height of 103 pc. Combining the fitting result of the Galactic stellar thin disk from similar data set \citep{Chen2017}, our results show that the dust disk is larger but thinner than stellar disk. The ratio of the scale-length of the Galactic dust disk to stellar disk is $h_d/h_s\sim 1.4$ while the ratio of scale-height is $z_d/z_s\sim 1/3$, which are in good agreement with the results of other nearby edge-on spirals \citep{Bianchi2007}. In our modeling, the vertical distance of the Sun to the dust disk plane is $23$ pc, which is in excellent agreement with that to the stellar plane \citep[ 25$\pm$5 pc][]{Ivezi2008}. That means the dust disk is coplanar with the stellar disk very well.

In our modeling, the radial distance of the Sun to Galactic center $R_{\odot}$ is fixed to 8.2 kpc \citep{Bland2016}. In principle, this parameter can be set free and then be constrained from data. However, we find that this parameter is highly degenerated with another model parameter, the central effective dust density $\rho_0\sigma$. The change of $R_{\odot}$ only changes the fitting result of $\rho_0\sigma$ and has negligible effects on other three structural parameters.

Besides the smooth disk component, we also have identified two dust substructures in solar neighborhood. One of the region, located at $150^{\circ}<l<200^{\circ}$, $-30^{\circ}<b<-5^{\circ}$ and extended from 0 to 500 pc, is overlapped with the Gould Belt well. Quantitatively, the combination of our best dust disk model with G09 Gould Belt model can reproduce the observed $E(B-V)$ growth curves for most of the sight lines in this region. The other substructure, region B, located at $140^{\circ}<l<165^{\circ}$, $0^{\circ}<b<15^{\circ}$, is associated with the Camelopardalis molecular clouds.

In our study, we have assumed that the dust extinction curve is uniform and so that the observed reddening $E(B-V)$ can be uniquely converted to the density of dust grains. The validation of our assumption relies on the fact that we only use the mean $E(B-V)$ values of large grids as tracers of the smooth dust component. Except solar neighborhood, the volume of the grids we used ranges from $\sim 10,000$ pc$^3$ to $\sim 0.02$ kpc$^3$ (dependent on heliocentric distance), which is much larger than typical molecular clouds. Therefore, an assumption of a uniform extinction curve on this scale is reasonable \citep{Schlafly2016}. On the other hand, a systematical change of extinction coefficients will only change the fitting result of $\rho_0\sigma$ and has no effect on other parameters. The statistical method we used not only benefits the estimation of dust extinction but also improves the accuracy of the distance estimation, which is quite uncertain for individual stars. Because of this, in Figure \ref{belt} and Figure \ref{badfit}, even we have increased the distance resolution from 250 pc to 100 pc, the statistical accuracy of the distance is still good enough to resolve the substructures of the dust distribution.

In our study, we have not considered any possible systematical bias in the $E(B-V)$ values of the catalog of Yuan et al. Any systematical bias, if exists, would also introduce bias into the distance estimation. Then, the dust structural parameters we have estimated would also be biased systematically. However, such a possibility is outside the scope of this study. 

\acknowledgments
LL thanks Professor Biwei Jiang, Jiawen Li and Li Yang for help of useful discussions. Guoshoujing Telescope (the Large Sky Area Multi-Object Fiber Spectroscopic Telescope LAMOST) is a National Major Scientific Project built by the Chinese Academy of Sciences. Funding for the project has been provided by the National Development and Reform Commission. LAMOST is operated and managed by the National Astronomical Observatories, Chinese Academy of Sciences. This work is supported by the ``973 Program'' 2014 CB845705/702 and National Natural Science Foundation of China (NSFC) with the Project Numbers  11573050,  11433003.


\end{CJK}
\end{document}